\documentclass[pre,showpacs,twocolumn,letter]{revtex4-2}
\usepackage[framemethod=tikz]{mdframed}
\usepackage{lipsum}
\usepackage{color}
\usepackage{epsfig}
\usepackage{amssymb,amsmath,amsthm}
\usepackage{esint}
\usepackage[colorlinks=true,linkcolor=blue,citecolor=blue]{hyperref}
\usepackage{graphicx}
\definecolor{red}{rgb}{1,0,0}

\begin{document}
\title{Statistics of a granular cluster ensemble at a liquid-solid-like phase transition}
\author{Enrique Navarro}
\email{enrique.navarro@ug.uchile.cl}
\author{Claudio Falc\'on}
\email{cfalcon@uchile.cl}
\affiliation{Departamento de F\'isica, Facultad de Ciencias F\'isicas y 
Matem\'aticas, Universidad de Chile, Casilla 487-3, Santiago, Chile}
\date{\today}

\begin{abstract}
We report on the construction of a granular network of particles to study the formation, evolution and statistical properties of clusters of particles developing at the vicinity of a liquid-solid-like phase transition within a vertically vibrated quasi two-dimensional granular system. Using the data of particle positions and local order from Castillo {\it et al}  [Phys. Rev. Lett. {\bf 109}, 095701 (2012)], we extract granular clusters taken as communities of the granular network via modularity optimization. Each one of these communities is a patch of particles with a very well defined local orientational order embedded within an array of other patches forming a complex cluster network. The distribution of cluster sizes and life-spans for the cluster network depend on the distance to the liquid-solid-like phase transition of the quasi two-dimensional granular system. Specifically, the cluster size distribution displays a scale-invariant behavior for at least a decade in cluster sizes, while cluster lifespans grow monotonically with  each cluster size. We believe this systematic community analysis for clustering in granular systems can serve to study and understand the spatio-temporal evolution of mesoscale structures in systems displaying out-of-equilibrium phase transitions.

 \end{abstract}
%\pacs{
% 47.15.gm, 04.70.−s, 04.80.Cc, 47.55.N
%}
\date{\today}

\maketitle

\section{Introduction}
Dry granular matter, {\it i.e.} a large collection of macroscopic particles interacting via dissipative collisions, can be driven into different phases (such as solid, liquid and gas-like ones) which depend on the energy injection-dissipation balance occurring within the system~\cite{AransonBook,JaegerRev,deGennesRev,AransonRev}. These out-of-equilibrium equilibrium phases display interesting transitions, which have been studied using very well-known theoretical tools relying on symmetry, dimensionality and conservation arguments~\cite{PhaseTransitionBook,HohenbergRev,BondBook}. These arguments allow a generic and universal way to characterize, in particular, the granular system's macroscopic evolution and properties usually linked to the large scale, slow modes that dominate the dynamics of the system. In this regard, the locality of interactions between grains are smeared out on the large scale dynamics. Thus, the local granular information (such as local force fluctuations and/or particle agglomerations) is lost within this modeling. Nevertheless, this local information is of paramount importance for the mechanical stability of granular matter when force chains are present~\cite{JaegerRev,NeddermanBook,Golhirsch2002,Snoeijer2004}, as well as for the description of defects in vibrated granular matter~\cite{Douady1989,Melo1995,Watanabe2008,Schindler2019}, specially in the case of structured granular systems (such as the case of nonisometric grains~\cite{Galanis2010,Velasco2017,Gonzalez-Pinto2018,Gonzalez-Pinto2019,Basurto2020,Diaz2020}). 

Recently, network science methodologies and techniques have been implemented in the study of granular systems~\cite{Richard1999,Hermann2005,Umbanhowar2007,Tordesillas2010,Ardanza2014,Tse2014,Diksman2018,Daniels2018,Porter2020} in order to understand the effect and importance of the local information (encoded into a granular network of forces, positions or bond orientational order) on the overall dynamics of the system. Quasi-two dimensional granular systems have been the main subject of study in this approach due to the direct accessibility to the information of each particle (such as position, velocity and force). It must be noticed that this feature has has already been exploited extensively to track out-of-equilibrium two-dimensional phase transitions~\cite{Urbach1998,Prevost2004,Reis2006,Castillo2012,Sun2016,Chaikin2022,Cates2023}. From the local information, different types of networks can be constructed, defined solely by the definition of nodes and their connections. In this paradigm, encoding the complex relations of the  granular system's particles in a rather simple networks can be extremely useful in the study of the granular system's static and/or dynamical properties, particularly the way it creates mesoscopic structures such as force chains~\cite{Daniels2015,Daniels2016} or clusters~\cite{Kou2018,Yi2020}.  

In this work, we study the formation, evolution and statistics of clusters of particles at the vicinity of a liquid-solid-like phase transition vía community detection by optimizing a quality function called modularity~\cite{Newman2006}. Experimental data of particle orientational and spatial order taken from Ref.~\cite{Castillo2012} is used to construct a network which encodes within its links the local orientational order of the granular system, enabling the use of the tools of network science within the framework of non-equilibrium phase transitions. To wit, we relate clusters of ordered particles of the non-equilibrium granular system to hard partitions of its respective network, which are computed using a null model for granular matter, validated via an entropic argument. 
\section{Bond-orientational network}

\begin{figure*}[t]
\includegraphics[width=1\textwidth]{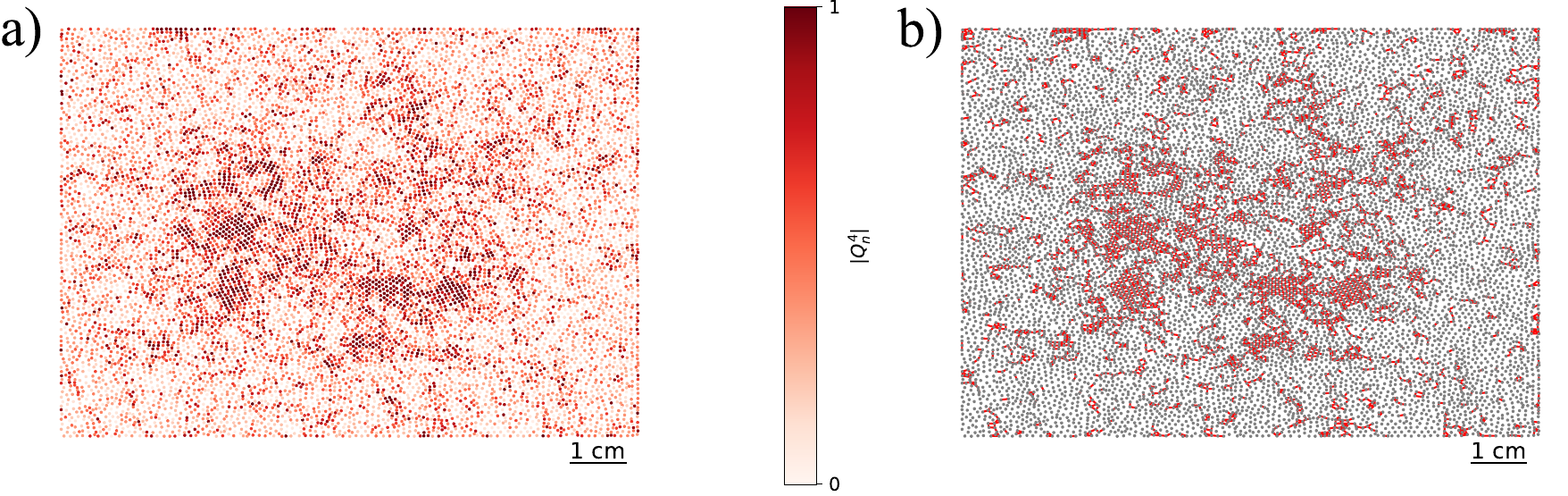}
\caption{Image of steel particles (1 mm in diameter) used to construct our clusters for $\Gamma=4.95$. a) $|Q_n^{4}|$ of each each particle. Color bar shows available values of $|Q_n^{4}|$. b) Links between particles used to construct the order network (in red) following Eq.~(\ref{Eq:AdjParticle}).}
\label{Fig:DataRaw}
\end{figure*}

A set of bond-orientational networks were constructed from data sets used in Ref.~\cite{Castillo2012}, where $\mathcal{N}=11504$ stainless steel spherical particles of diameter $d=1$ mm are confined in a box of lateral dimensions $100 d \times 100 d$ and a vertical dimension of  $1.94 d$. The box is vibrated vertically with in a sinusoidal acceleration $a(t)=\Gamma g \cos{(\omega t)}$, where $g$ is gravity, $1<\Gamma<6$ and $\omega=200\pi$ rad/s.  3000 images of the in-plane motion of the particle ensemble were acquired at 10 fps and used to detect them with subpixel accuracy using a particle tracking algorithm. From each image, the position of each particle in the horizontal plane $\vec{r}_n$ and the fourfold bond-orientational order parameter per particle 
\begin{equation}
Q^{4}_n=\frac{1}{\mathcal{M}_n} \sum_{m\in\mathcal{M}_n} e^{i 4\alpha
_{nm}}
\end{equation}
were calculated. Here $\mathcal{M}_n$ is set of nearest neighbors of particle $n$ computed via a Voronoi partition and $\alpha_{nm}$ is the angle between the neighbor $m$ of particle $n$ and a given axis. From this point on, we will set $Q_n^{4}\equiv Q_n$, as we will only consider four-fold orientational order.

For a given $\Gamma$, a (simple) network represented by an adjacency matrix
\begin{equation}
A_{nm}^{(p)}=\begin{cases}
    1 & \mbox{if  } \,  |Q_n| ,|Q_m|\geq \langle|Q|\rangle\, \, \mbox{and  } \,  m\in\mathcal{M}_n\\
    0 & \mbox{otherwise  } \,\\
    \end{cases}
\label{Eq:AdjParticle},
\end{equation}
is computed for each image using $\langle|Q|\rangle$ as a threshold for the solid-liquid transition~\cite{Castillo2012}. The results presented here display no change when $\langle|Q|\rangle$  is varied within $\pm10\%$. Non-directionality for the network connections is enforced by the symmetric nature of $A_{nm}^{(p)}$. Tadpoles are directly removed from the adjacency matrix as no self-links are available ($A_{nn}\equiv0$ for all $n$)~\cite{NewmanBook}.

The binary adjacency matrix computed above encodes the connections (edges) between particles (nodes) $n$ and $m$ with a large local orientational order. In Fig~\ref{Fig:DataRaw}a) a snapshot of particles at a given $\Gamma$ is presented displaying the local value of  $|Q_4|$ as a color map over each particle, showing that they arrange themselves in cluster of similar $|Q_4|$ values. Using Eq.(\ref{Eq:AdjParticle}), the connections given by the adjacency matrix are shown with red lines between particles in Fig.~\ref{Fig:DataRaw}b)  We relate the particle clusters displayed in Fig~\ref{Fig:DataRaw}a) via the array of connections displayed in Fig~\ref{Fig:DataRaw}b) to the network's communities~\cite{NewmanBook}, which are indivisible subgroups within the network (what is called a hard partition). In a similar fashion one can encode the strength of the connections between nodes vía a weighted adjacency matrix $W_{nm}^{(p)}$ between nodes $n$ and $m$, which in our case can be readily defined as $W_{nm}^{(p)}=|Q_n| |Q_m| A_{nm}^{(p)}$.

In our work, the clusters are found by maximizing a certain quality function, the network's {\it modularity} $\mathcal{Q}$,
\begin{equation}
\mathcal{Q}=\sum_{n,m\in\mathcal{N} } (W_{nm}^{(p)}-\gamma P_{nm})\delta_{C_n,C_m}
\label{Eq:Modularity}
\end{equation}
where node $n$ resides in community $C_n$ and node $m$ resides in community $C_m$, $\gamma$ 
is called the resolution parameter, $P_{nm}$ is a matrix term stemming from a null model~\cite{Sarzynska2016} for the edge distribution within the network, and $\delta_{x,y}$ is Kronecker's delta function. Modularity has been proposed as a direct way to measure and quantify the community structure within large networks~\cite{Newman2006}, where communities (or modules) are nodes that have a larger ammount of non-zero connections among themselves than with the rest of the network's nodes~\cite{Barthelemey2011}. One can understand $\gamma$ as the ratio between the spatial densities of two communities in an optimal hard partition following the null model selection. When $\gamma<1$ the hard partition tends to favor the selection of larger communities rather than smaller ones, and vice versa~\cite{Sarzynska2016}. We have set $\gamma$=1 as we try to find the cluster dynamics at the solid-liquid-like transition where the system tends to agglomerate into clusters with a large span in scales (ideally, in a scale invariant way). 
\begin{figure*}[t!]
\includegraphics[width=0.32\textwidth]{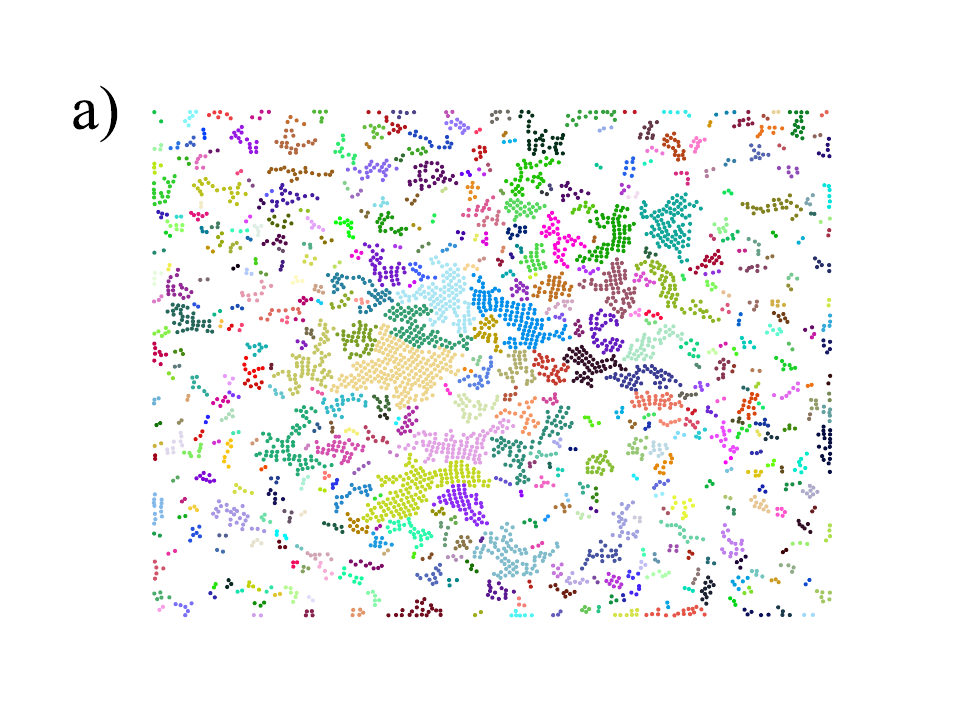}
\includegraphics[width=0.32\textwidth]{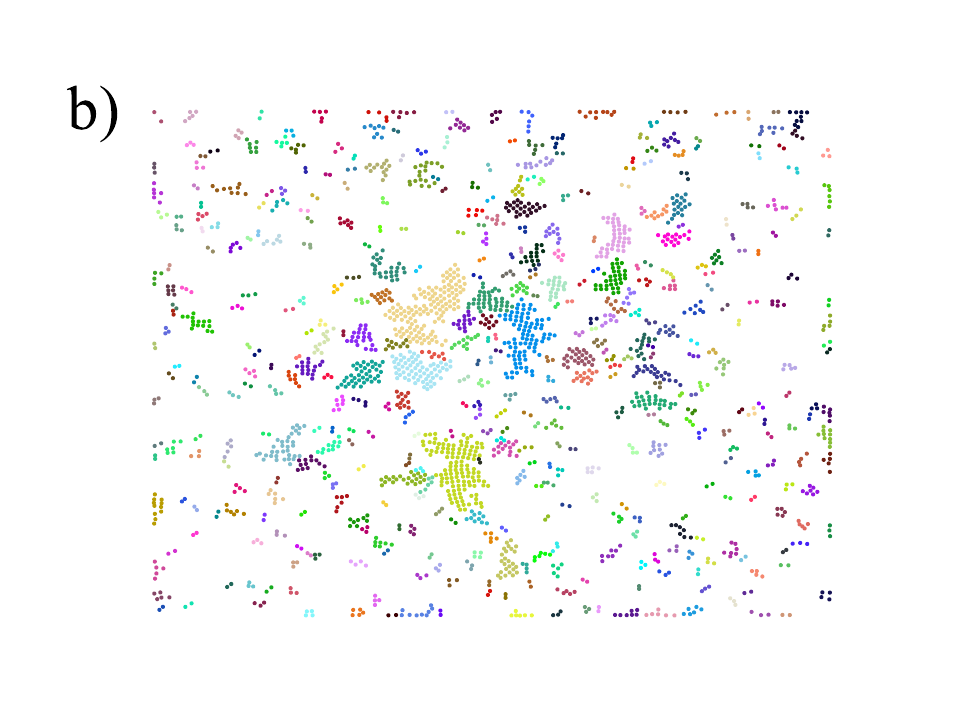}
\includegraphics[width=0.32\textwidth]{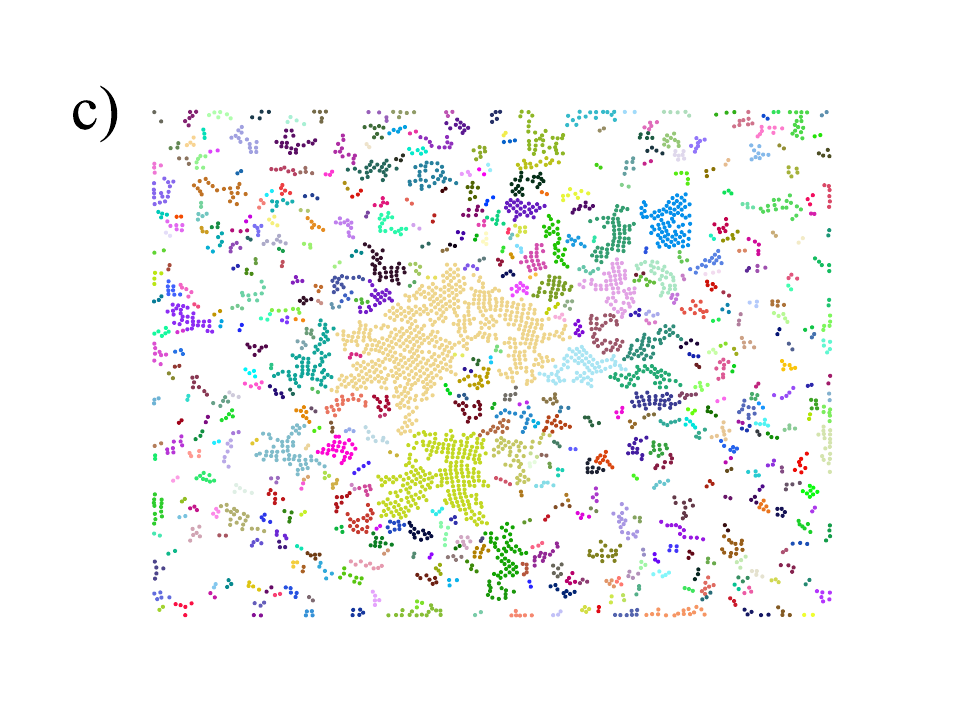}
\caption{Examples of communities found from optimizing  $\mathcal{Q}$ using (a) the Newman-Girvan, (b) the geographic, and (c) our modified null model for $\gamma=1$ and $\Gamma=4.95$. Patches of particles with the same color represent communities. Particles with no connections (edges) as well as singleton clusters are not shown as they do not contribute to the adjacency matrix.}
\label{Fig:NullModel}
\end{figure*}

We have tested  3 different null models to track the structure of particle clusters: the Newman-Girvan null model~\cite{Newman2004a}, the geographic null model~\cite{Daniels2018} and a modification of the former one, adapted to the observed data. The Newman-Girvan null model, which has been extensively used in community detection~\cite{Sarzynska2016} is based on edges which are placed at random on each node. The randomness of this edge configuration is quantified by the degree $k_n$ of each of the network's nodes ({\it i.e.} the number of edges connecting it)~\cite{NewmanBook}.  To wit, $P_{nm}=k_n k_m/\kappa$ where $\kappa=2\sum_{n} k_n=2\mathcal{N}\langle k\rangle$ is the total number of edges of the network and $\langle k\rangle$ is the average number of edges per node on the network. For granular matter, the hypothesis of random connections for each and every node does not correspond to the reality of the local granular network connectivity. To correct this hypothesis, Daniels and co-workers~\cite{Daniels2018} proposed a geographic null model, where nodes represent particles and edges represent, for instance, forces between them, which are encoded into $P_{nm}=\langle f\rangle A_{nm}^{(p)}$ with $\langle f\rangle$ the mean inter-particle force. This null model displays communities which follow the local force chains of the granular material. 
\begin{figure}[b]
\includegraphics[width=0.5\textwidth]{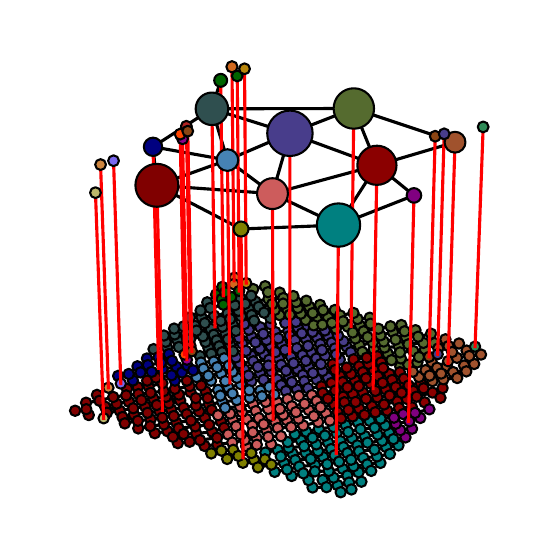}
\caption{A given cluster of particles (in the bottom plane) is represented by a cluster node (in the upper plane) for $\Gamma=4.83$. The red lines show the correspondance between the two networks. The black lines (in the upper plane) show the edges of the cluster network which represent the amount of particles at the edges between particle clusters (in the bottom plane). }
\label{Fig:Modularity}
\end{figure}

These two null models display very different community structures when we maximize $\mathcal{Q}$ for a given $\Gamma$ and $\gamma=1$, as shown in Fig.~\ref{Fig:NullModel}. Here only connected particles are depicted, which contribute to non-zero terms to the adjacency matrix. Using the Newman-Girvan null model,(cf. Fig.~\ref{Fig:NullModel}a)) the cluster array found from the $\mathcal{Q}$ optimization displays clusters with sizes that are exponentially distributed with several inter-cluster connections, much more than the ones found using the geographic null model (cf. Fig.~\ref{Fig:NullModel}b)). A particular characteristic of the cluster array arising from the use of the geographic model is the appearance of single particle clusters within clusters of 4 to 10 particles, which develop as a consequence of the nature of a model constructed to follow force chains in compacted granular matter~\cite{Daniels2018}. Following this feature, we propose a modified model to take into account the quality of the ordering between neighboring particles. We propose the null model $P_{nm}=\beta \langle |Q|\rangle^{2} A_{nm}^{(p)}/2$ following the geographic model approach, where $\beta$ is the fraction of ordered particles within the system (see the Appendix for an entropic justification of this null model).
\begin{figure*}[t]
\includegraphics[width=1\textwidth]{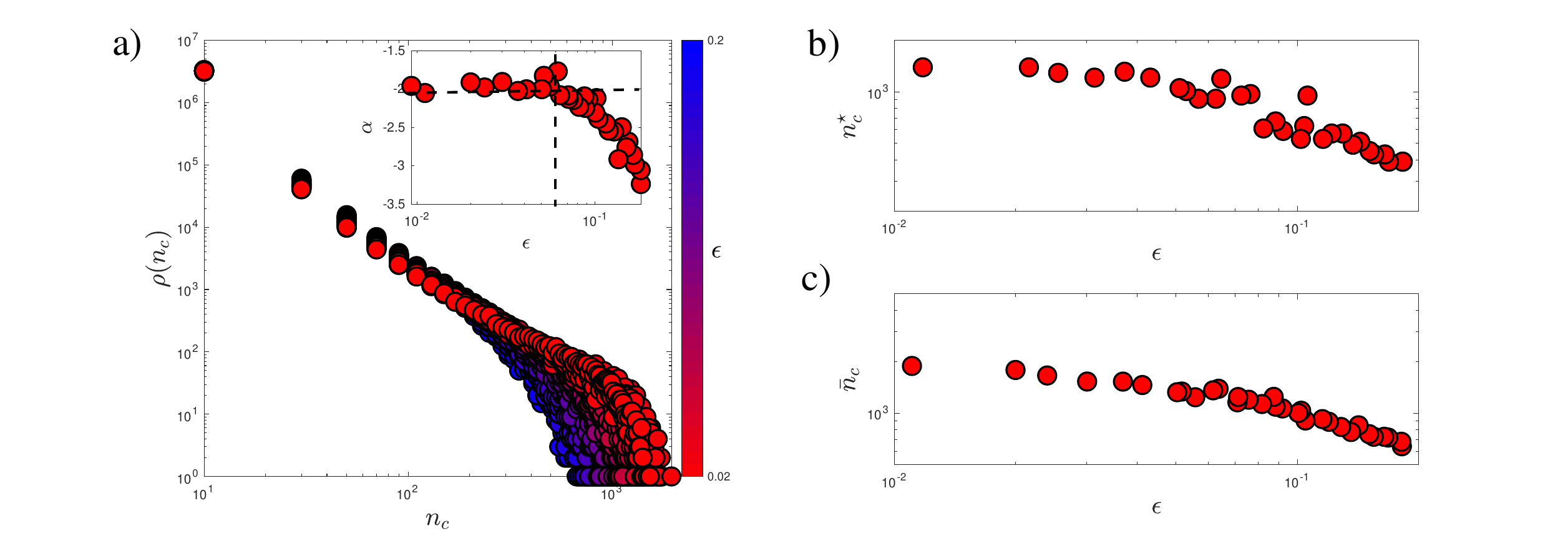}
\caption{a) Distribution of cluster sizes $\rho(n_c)$ vs $n_c$ for $0.02<\epsilon<0.2$. Inset: Best fit power-law exponent $\alpha$ as a function of $\epsilon$ for $N_c\in\{50,500\} $. Horizontal dashed line shows the limiting value of $\alpha$ as a function of $\epsilon$. Vertical dashed line shows the larges value of $\epsilon$ from which $\alpha saturates$. Color bar shows values of $\epsilon$. b) Cluster cut-off $n_c^{\star}$ as a function of $\epsilon$. c) the Largest cluster distance $\bar{n}_c$ as a function of $\epsilon$. }
\label{Fig:RhoNc}
\end{figure*}
To find algorithmically these communities we optimize the community partition by maximizing $\mathcal{Q}$ using a local greedy maximization algorithm~\cite{{Newman2004b,LouvainGreedy2008}} in  such a way that the total edge weight within the communities is as large as possible with respect to a chosen null model~\cite{Sarzynska2016}. The optimally computed community partition is then recomputed a number of times (chosen heuristically~\cite{Fortunato2018}) in order to assure a certain convergence as the maximization process is NP-hard~\cite{NPBrandes2008}. In our maximization process we have repeated the calculation at least 20 times per images by permuting the nodes, finding the same clusters. 
\section{Cluster network }
After the above optimization process, we find a set of communities (clusters) $C_\alpha$ of particles which are connected by intra-communal edges ($A_{nm}^{(p)}$ for $n,m$ $\in C_\alpha$). These clusters are also connected between them via inter-communal edges ($A_{nm}^{(p)}$ for $n$ $\in C_\alpha$, $m\in C_\beta$ and $\alpha\neq\beta$) thus forming a new coarse-grained network. We construct a new adjacency matrix for the (much smaller) cluster network
\begin{equation}
A_{\alpha \beta}^{(C)}=
\begin{cases}
    1 & \mbox{if  } \,  A_{nm}^{(p)}=1 \, \, \mbox{for any } \,  n\in C_\alpha, m\in C_\beta,\\
    0 & \mbox{otherwise.  } \,\\
    \end{cases}
% \sum_{n\in C_\alpha, m\in C_\beta }A_{nm}^{(p)}  
\label{Eq:AdjCluster}
\end{equation}
which is used to describe the cluster evolution as a function of $\Gamma$ near the solid-liquid transition occurring at $\Gamma_c =5.09\pm 0.07$~\cite{Castillo2012}. This means that the inter-communal edge  between two clusters is nonzero if there is at least one particle shared at the edges of these clusters.  A representation of this clusterisation process is shown in Fig.~\ref{Fig:Modularity} using a small region of an image at $\Gamma=4.95$. The detected clusters of particles, represented by large nodes with sizes scaling with the number of particles per cluster, display inter-cluster connections (black lines) which represents that there are particles at the edges between clusters.As reported by Castillo~\cite{Castillo2012}, the granular system displays an increasing global orientational order as $\langle |Q|\rangle$ grows linearly with $\Gamma<\Gamma_c$. As this occurs, particles arrange themselves into clusters with a larger and larger number of particles that adjoin other clusters via shared particles at their edges. Using the normalized acceleration $0.02<\epsilon=(\Gamma_c-\Gamma)/\Gamma_c <0.2$, we compute the distribution of sizes $n_{c}$, 
and lifespan $\tau_c$ of the clusters from these network sets as a function of $\epsilon$. 

\subsection{Size and lifespan properties of the cluster network}
The distribution of cluster sizes $\rho (n_c)$ using our modified  null model is shown in Fig.~\ref{Fig:RhoNc} for different values of $\epsilon$. It displays a power-law behavior for $\rho(n_c)$ as a function of $n_c$, $\rho(n_c)\sim n_c^{\alpha}$ for at least a decade in cluster sizes between 20 to 200 particles per cluster. The best fit values for $\alpha$ within the above range are depicted in Fig.~\ref{Fig:RhoNc} a) (inset), showing a limiting behavior for low $\epsilon$. This power law behavior saturates displaying a cut-off for cluster sizes larger than a given value $n_c^{\star}(\epsilon)$ which is larger and larger as we approach the liquid-solid-like transition. Using a simple exponential correction $\rho(n_c)\sim n_c^{\alpha}\times \text{exp}(-n_c/n_c^{\star})$, we compute $n_c^{\star}=n_c^{\star}(\epsilon)$ which is displayed in Fig.~\ref{Fig:RhoNc} b). Similarly, one can compute the largest cluster distance found for each value of $\epsilon$, $\bar{n}{_c}=\langle n_c ^{2} \rangle$ which is shown in Fig.~\ref{Fig:RhoNc} c). As in the case of $n_c^{\star}$, $\bar{n}_c$ decreases with $\epsilon$. One can probe a power-law behavior for both $n_c^{\star}$ and  $\bar{n}_c$ in $\epsilon$ in the range where $\alpha$ converges towards -2.0, as it is depicted in Fig.~\ref{Fig:RhoNc} a) (inset). Within this range, $n_c^{\star}\sim \epsilon^{\zeta^{\star}}$  with $\zeta^{\star}=0.23+0.05$ and  $\bar{n}_c\sim \epsilon^{\bar{\zeta}}$ with $\bar{\zeta}=0.21+0.05$. One might adscribe the above power-law behavior for cluster size distribution to percolation-related problems~\cite{Stauffer1979,Isichenko1992,Saberi2015,Li2021} as it displays the predicted exponent for our cluster distribution. Recently, a couple of different second-order phase transitions (one of activity and one of orientational order) have been found to appear in a system of two-dimensional sheared granular discs at the same critical value of the control parameter~\cite{Chaikin2022}, which might be the case of the data from Ref.~\cite{Castillo2012}. We have checked that for our cluster network, this is not the case, as neither $n_c^{\star}$ nor $\bar{n}_c$ display the critical behavior expected in percolation-related problems as a function of $\epsilon$ within the experimental range of the present data. 

\begin{figure*}[t]
\includegraphics[width=0.30\textwidth]{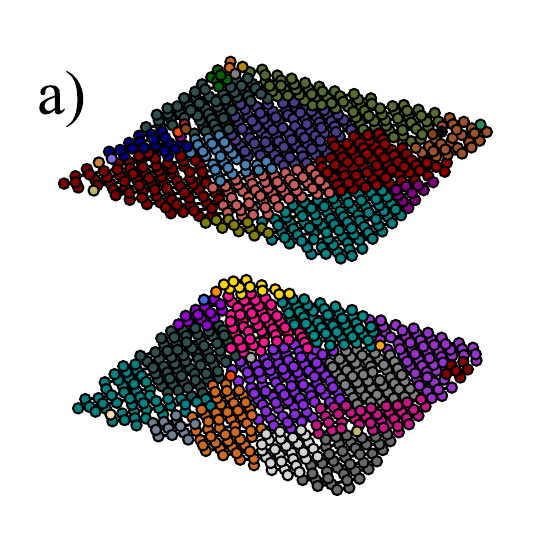}
\includegraphics[width=0.30\textwidth]{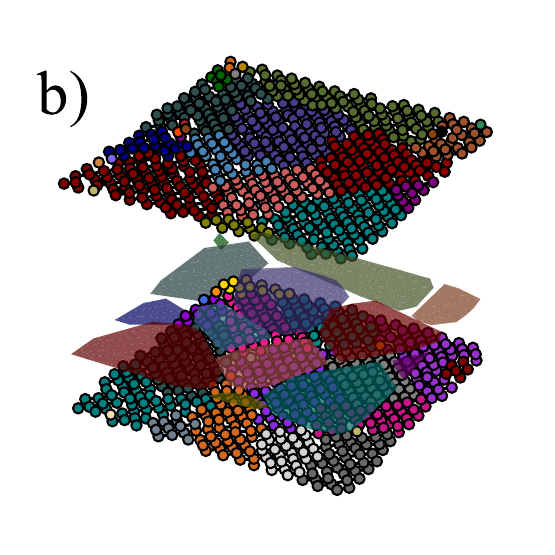}
\includegraphics[width=0.30\textwidth]{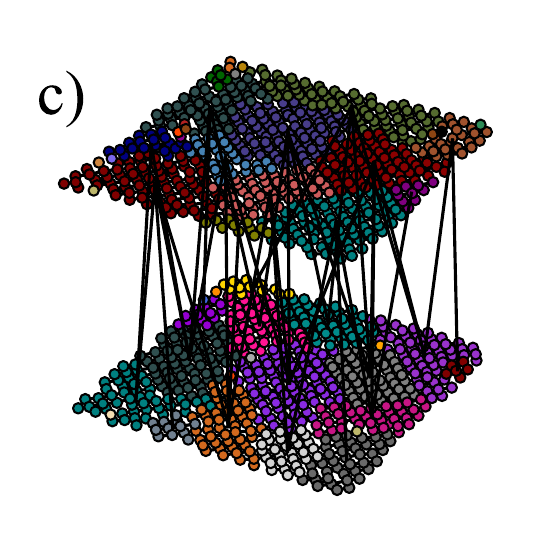}
\includegraphics[width=0.30\textwidth]{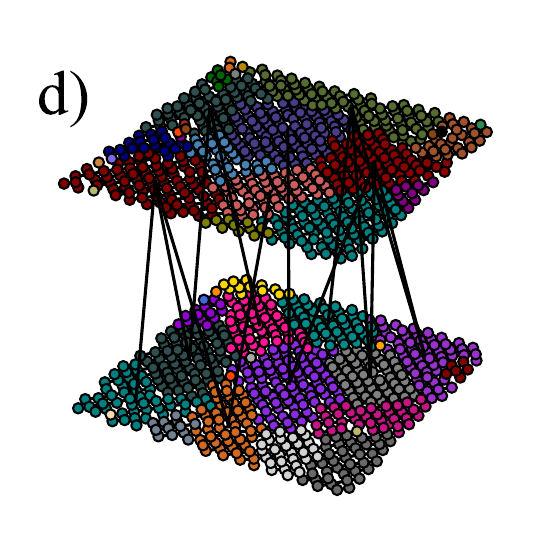}
\includegraphics[width=0.30\textwidth]{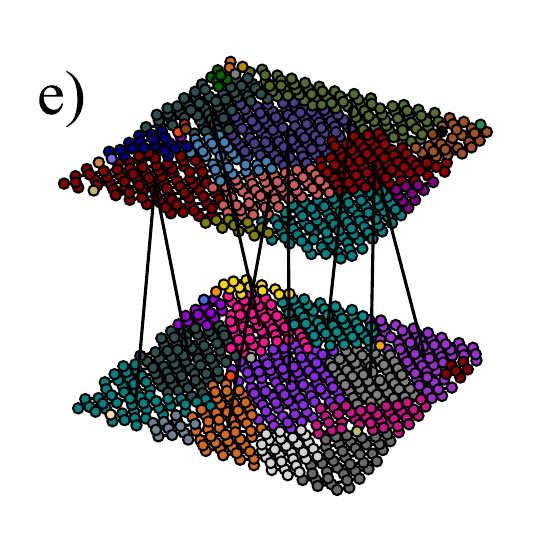}
\includegraphics[width=0.30\textwidth]{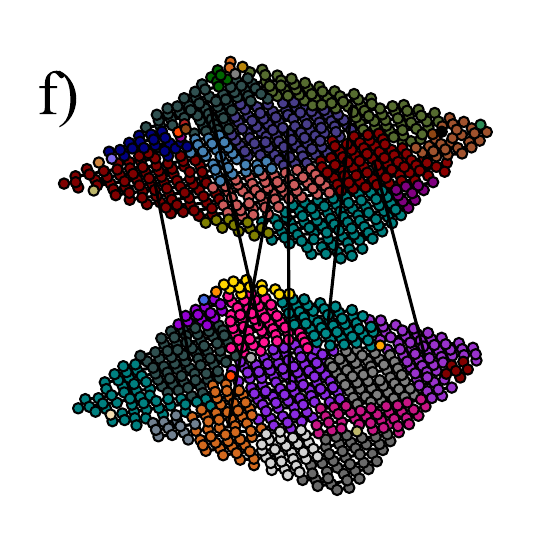}
\caption{Temporal cluster evolution procedure for two consecutive images. (a) Cluster network detection for both images. (b) Area projection of one cluster network onto the other. (c) Link creation between clusters. (d) Link trimming by weight. (e) Backward link trimming. (f) Forward link trimming.}
\label{Fig:TemporalCluster}
\end{figure*}

Using these cluster network sets, we track the temporal cluster evolution for a given $\epsilon$ to compute the average lifespan of clusters of size $n_c$. The way we track the cluster temporal evolution is depicted in Fig.~\ref{Fig:TemporalCluster}. The process starts by tracking the cluster networks within two consecutive images $j$ and $j+1$ (Fig.~\ref{Fig:TemporalCluster}a)). The area of each cluster from image $j$ is projected onto image $j+1$  (Fig.~\ref{Fig:TemporalCluster}b)) and a link is created between a cluster $C_n^{j}$ of image $j$ and a cluster $C_m^{j+1}$ of image $j+1$ if nodes from $C_m^{j+1}$ correspond to the projected area of $C_n^{j}$ (Fig.~\ref{Fig:TemporalCluster}c)) .This link has a directed weight $p^{(f)}_{nm}$ equal to the ratio of the number of nodes of $C_n^{j}$  and $C_m^{j+1}$.  In the same fashion there is a directed weight $p^{(b)}_{mn}$  equal to the ratio of the number of nodes of $C_m^{j+1}$  and $C_n^{j}$. We keep only the links with $(p^{(b)}_{nm},p^{(f)}_{nm})>0.5$, which means that both clusters share at least half of the nodes (Fig.~\ref{Fig:TemporalCluster}d)). As $C_m^{j+1}$ can be linked with more than one cluster $C_n^{j}$, we restrict the linkage between clusters by keeping the one with the largest $p^{(b)}_{mn}$ (Fig.~\ref{Fig:TemporalCluster}e)). The same procedure is done then forwards in time with $p^{(f)}_{nm}$, to link only one cluster $C_n^{j}$ to only cluster $C_m^{j+1}$ (Fig.~\ref{Fig:TemporalCluster}f)). Following this link path for each cluster from image $j=1$ to  $j=\mathcal{N}$ we can track its lifespan as the number of links it holds until the cluster is no longer traceable. 

\begin{figure}[t!]
\includegraphics[width=0.5\textwidth]{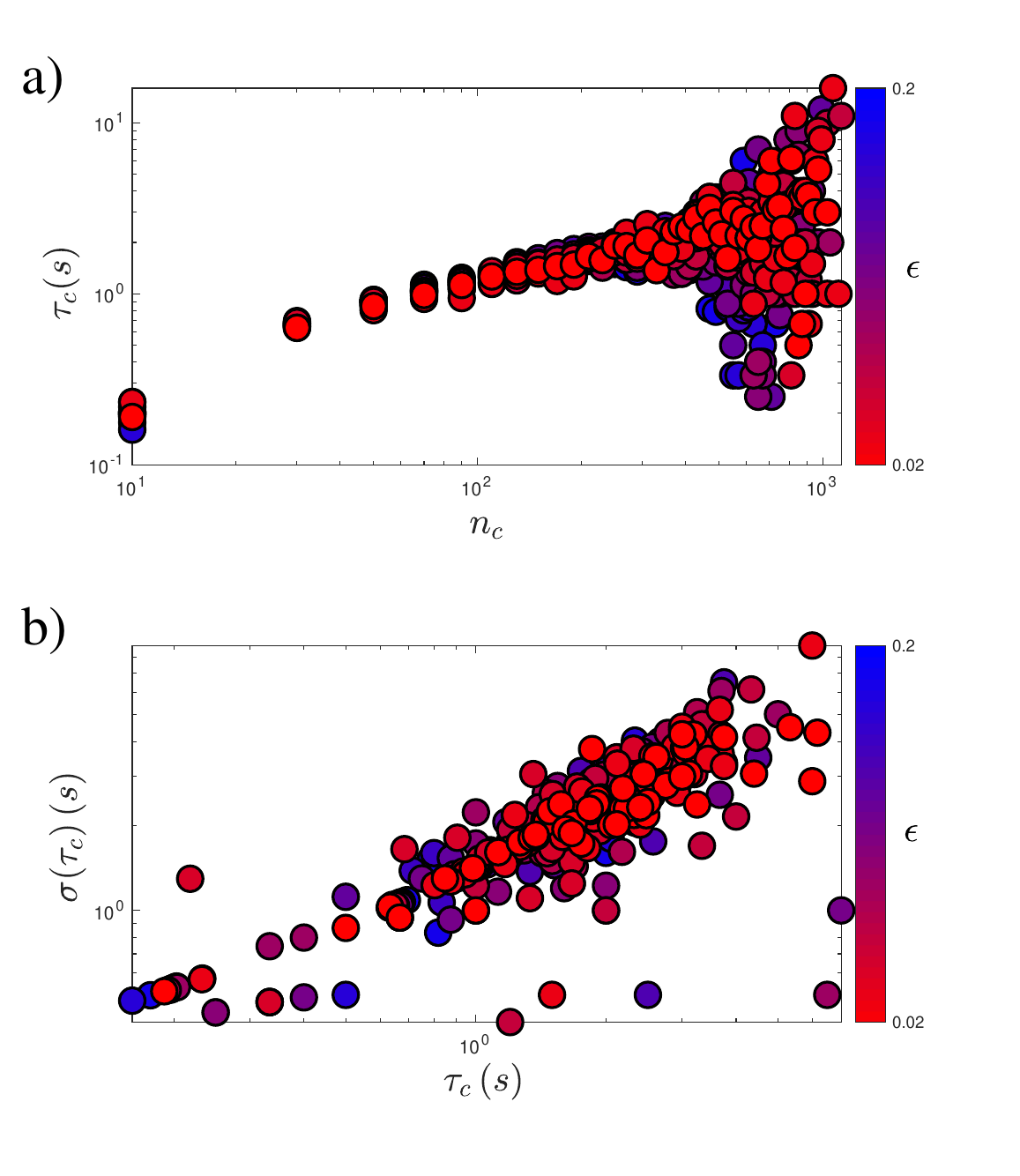}
\caption{a) Average lifespan $\tau_c$ as a function of $N_c$ as a function $\epsilon$. b) Standard deviation of the average lifespan $\sigma(\tau_c)$ vs the average lifespan $\tau_c$. }
\label{Fig:Tau}
\end{figure}

From the above procedure, the average lifespan $\tau_c=\tau(n_c)$ as a function of $n_c$ is shown in Fig.~\ref{Fig:Tau} using the experimental sample frequency of 10 Hz. Clusters with sizes in $(n_c-\Delta n_c,n_c+\Delta n_c)$ are binned together using $\Delta n_c=$20 which is the average size of the small clusters found around the edges of the images (see Fig.~\ref{Fig:DataRaw}). For a fixed $\epsilon$, $\tau_c$ grows monotonically with $n_c\in\{50,500\} $ as a power law with exponent $\eta=0.24\pm0.05$  as a best fit slope. Clusters with larger $n_c$ are sustained with a larger $\tau_c$, displaying a large dispersion as $n_c$ is of the order of 1000 particles for all values of $\epsilon$ as shown in Fig.~\ref{Fig:Tau}a). Lifespan fluctuations $\sigma(\tau_c)$ are depicted in Fig.~\ref{Fig:Tau}b) as a function of $n_c$, which grow (on average) monotonically with $\tau_c$. A power law can be fitted as $\sigma(\tau_c)\sim\tau_c^{\mu}$, with $\mu=0.55\pm0.05$ as the best fit slope. This shows that as clusters grow in size and thus $\tau_c$ increases, $\sigma(\tau_c)/\tau_c\sim\tau _c^{-1/2}$ decreases, which means that $\tau_c$ can be used a proper time scale to describe the slow dynamics of granular cluster evolution. It is important to notice that this findings are also observed when $(p_{nm}^{(f)},p_{nm}^{(b)})$ are reduced from 0.5 to 0.1, albeit larger larger fluctuations in $\tau_c$ are found. 

\section{Conclusions}
The results presented above display a community detection scheme via a $\mathcal{Q}$ optimization which allows the computation of a cluster ensemble of ordered particles in a quasi-two dimensional vibrated granular system close to a solid-to-liquid-like phase transition. These clusters are constructed as hard partitions using an entropic null model in $\mathcal{Q}$ which takes into account both the bond-orientational  and the spatial order of the granular system. From the computed cluster ensemble as we increase the normalized acceleration $\epsilon=(\Gamma_c-\Gamma)/\Gamma_c$, we track the cluster size distribution $\rho(n_c)$ and the cluster mean lifespan $\tau_c$. The cluster size distribution displays a power-law dependence on $n_c$ with an exponent close to -2.0 which  independent of $\epsilon$ in the range $n_c\in\{50,500\}$, and an exponential cut-off with a slope equal to $1/n^{\star}_c(\epsilon)$ which increases with  $\epsilon$. The mean lifespan $\tau_c$ of clusters with size $n_c$ increases with a power-law as a function $n_c$ with an exponent close to $1/4$ for $n_c\in\{50,500\}$. For larger values of $n_c$, large fluctuations of $\tau_c$ are observed.

We believe that this community analysis vía $\mathcal{Q}$ optimization for cluster detection in granular systems can be of use to study and understand the spatio-temporal evolution of mesoscale structures in systems, specially ones displaying out-of-equilibrium phase transitions. Furthermore, the application of an entropic null model in the $\mathcal{Q}$ optimization scheme enables a systematic computation of mesoscale structures in out-of-equilibrium granular systems and their dynamics without the necessity of defining ad-hoc parameter values~\cite{Daniels2018}.  It is our hope that this scheme will be used on other quasi-two dimensional granular systems to study, compare and contrast the dynamics of their mesoscale structures as phase transitions develop in such systems. 

\section{Acknowledgements}

The authors wish to thank G. Castillo, N. Mujica and R. Soto for kindly allowing access to the database used in Ref.~\cite{Castillo2012}. This work was partially supported by FONDECYT Regular Grants 1190005 and 1210656.

\appendix
%\counterwithin{figure}{section}
%\counterwithin{table}{section}
\section{A Null model for ordered phases in granular matter}

The null model presented in the present work to find communities was found by an entropic maximization scheme. We start by considering a multiplex network~\cite{BiaconiBook,Biaconi2013,Biaconi2014}, which means that two nodes within one network can belong also to a different one at the same time. In our case we will consider a multiplex formed by $N$ labeled nodes from $n=1,...,N$ and $M$ layers represented by a $\vec{G}=(G^{(1)},...,G^{(M)})$, where $G^{(\alpha)}$ indicates the set of all possible networks at layer $\alpha=1,...,M$ of the multiplex. Nodes can be connected by specifying an adjacency matrix $A_{n,m}^{\alpha}$ (as defined in the text) which can be weighted, depending on the type of network under study.  

With this in mind, we task ourselves to find the ensemble properties of the multiplex by specifying the probability $P(\vec{G})$ for every possible multiplex. For a given set of multiplex probabilities $P(\vec{G})$ we can compute the entropy of the multiplex ensemble 
\begin{equation}
\mathcal{S}=-\sum_{\vec{G}}P(\vec{G})\log{(P(\vec{G}))},
\label{A:Entropy}
\end{equation}
which is proportional to logarithm of the number of possible multiplexes of the ensemble. As in statistical mechanics, we can find an ensemble of multiplexes that maximize $\mathcal{S}$, subjected to a given set of restrictions (a Gibbs probability ensemble). If we assume that the layers of the multiplex are uncorrelated, the probability of the multiplex can be written as 
\begin{equation}
P(\vec{G})=\prod_{\alpha=1}^{M}P(G^{\alpha})
\label{A:MultProb}
\end{equation}
which simplifies greatly our task. 

We now specify our problem within this framework. First we will set $M=2$ and $\alpha=o,s$. One layer, called the orientational layer ($o$), connects pairs of nodes $(n,m)$ with intralayer weights $w^{(o)}_{n,m}=|Q_n||Q_m|$ as long as $|Q_n|,|Q_m|>\langle |Q|\rangle$. A second layer, called the spatial layer ($s$), connects pairs of nodes $(k,l)$ with intralayer weights $w^{(s)}_{k,l}=f(d_{k,l})$ where $d_{k,l}=|\vec{r}_n - \vec{r}_m|$ is the spatial distance (measured in number of diameters $d$) between nodes and $f(x)$ is a scalar function that goes to zero with increasing $x$. %As one might suspect, while the orientational layer spans the orientational properties calculated over each node $n$ encoded on $Q_n$, the spatial layer spans the relative distances between particles encoded on $d_{k,l}$.
Note that in the orientational layer any pair of nodes $(n,m)$ that satisfy the thresholding scheme for $|Q_n|$ and $|Q_m|$ is connected. In this approximation, local orientational order does not couple directly with spatial order~\cite{Castillo2012}, we will assume for simplicity that $\langle w^{(o)}_{n,m}w^{(s)}_{n,m}\rangle=\langle w^{(o)}_{n,m}\rangle \langle w^{(s)}_{n,m}\rangle$, {\it i.e.}, we neglect the overlap between layers~\cite{Biaconi2014}. This means that the probability of finding a link between nodes $(n,m)$ in both layers of the multiplex is simply the multiplication of the intralayer link probabilities for these nodes, following Eq.(\ref{A:MultProb}). 

In a canonical multiplex ensemble for our data, the set of multiplexes $\vec{G}=(G^{(o)},G^{(s)})$ satisfy constraints on average, informed by the dynamics of the granular layer both in the orientational layer and in the spatial one. In this case, for the orientational layer we set the global constrain 
\begin{eqnarray}
L^{(o)}&=&\sum_{\vec{G}}P(\vec{G})\sum_{n<m}^{N}w_{n,m}^{(o)}\\ \nonumber
%&=&\sum_{n<m}|Q_n||Q_m| \\ \nonumber
&=& \beta \frac{N(N-1)}{2}\langle |Q^{\star}|\rangle ^{2}
\label{A:ConstraintL}
\end{eqnarray}
which can be understood as setting on average a fraction $\beta$ of nodes (particles) within the system with a fourfold bond-orientational order parameter per particle $\langle |Q^{\star}|\rangle$. For the spatial layer the constraint
\begin{equation}
0=\sum_{\vec{G}}P(\vec{G})\sum_{n<m}^{N}\Theta(w_{n,m}^{(s)})(\Theta(d_{nm}-d))
\label{A:ConstraintDelta}
\end{equation}
sets in average the spatial interaction of particles only the nearest neighbors for each particle. Here $\Theta(x)$ is the Heaviside function~\cite{ArfkenBook}. Thus, maximizing $\mathcal{S}$ from Eq.(\ref{A:Entropy}) constraint to Eqs.(\ref{A:ConstraintL})-(\ref{A:ConstraintDelta}) gives 
\begin{equation}
P(\vec{G})=\frac{e^{ \sum_{n<m}^{N}-\Lambda w_{n,m}^{(o)}- \Delta\Theta(w_{n,m}^{(s)})(\Theta(d_{nm}-d))}}{\mathcal{Z}}
\end{equation}
where $\mathcal{Z}$ is the partition function of our system and $(\Lambda,\Delta)$ are Lagrange multiplier enforcing the orientational ($\Lambda$) and spatial ($\Delta$) constraints of the system, respectively. As the layers are assumed to be uncorrelated, we calculate $\mathcal{Z}=\mathcal{Z}^{(o)}\times \mathcal{Z}^{(s)}$ with 
\begin{eqnarray}
\mathcal{Z}^{(o)}&=&\sum_{G^{(o)}} e^{- \Lambda \sum_{n<m}^{N}w_{n,m}^{(o)}}\nonumber\\
&=&\prod_{n<m}\sum_{\{w_{n,m}^{(o)}\}}e^{-\Lambda w_{n,m}^{(o)} }\nonumber\\
&=&\prod_{n<m}(1-e^{-\Lambda})^{-1}
\label{A:Zetao}
\end{eqnarray}
and 
\begin{eqnarray}
\mathcal{Z}^{(s)}&=&\sum_{G^{(s)}} e^{- \Delta  \sum_{n<m}^{N}\Theta(w_{n,m}^{(s)})(\Theta(d_{nm}-d))}  \nonumber\\
&=&\prod_{n<m}\sum_{\{\Theta(w_{n,m}^{(s)})\}=0,1} e^{-\Delta \Theta(w_{n,m}^{(s)})(\Theta(d_{nm}-d)) }\nonumber\\
&=&\prod_{n<m}(1+e^{-\Delta(\Theta(d_{nm}-d))}).
\label{A:Zetas}
\end{eqnarray}

It is now straightforward to compute the link probability ({\it i.e.}, the null model) for our mulitplex between nodes $(n,m)$ from the product partition function $P_{nm}$. The link probability between two nodes $P_{nm}^{(o)}$ for the orientational layer can be computed from the weight probability $\pi_{nm}(w)$ of the link for a given weight $w^{(s)}_{nm}=w$ 
\begin{eqnarray}
\pi_{nm}(w)&=& \sum_{\vec{G} }P(\vec{G})\delta(w^{(o)}_{nm}=w)\nonumber\\
&=& e^{-\Lambda w}\times(1-e^{-\Lambda}),
\end{eqnarray}
as $P_{nm}^{(o)}=1-\pi_{nm}(w=0)=e^{-\Lambda}$. Using Eq.(\ref{A:ConstraintL}) and Eq.(\ref{A:Zetao}), we link $\Lambda$ and the restriction via 
\begin{eqnarray}
L^{(o)}&=&-\frac{\partial \log{(\mathcal{Z}^{(o)})}}{\partial \Lambda}=\sum_{n<m} \frac{e^{-\Lambda} }{1- e^{-\Lambda} } \nonumber \\
&=&\frac{N(N-1)}{2} \frac{e^{-\Lambda} }{1- e^{-\Lambda} } \nonumber\\
&=&\beta \frac{N(N-1)}{2} \langle |Q^{\star}|\rangle^{2}, 
\end{eqnarray}
which sets $e^{-\Lambda}=\beta  \langle |Q^{\star}|\rangle^{2}/(1+ \beta \langle |Q^{\star}|\rangle^{2})=P_{nm}^{(o)}$. The link probability between two nodes $p_{nm}^{(s)}$ in the spatial layer can be computed similarly imposing that the weigh of the link t is larger than zero, {\it i.e.}, 
\begin{equation}
P_{nm}^{(s)}= \sum_{\vec{G} }P(\vec{G})\delta(w^{(s)}_{nm}>0)= \frac{e^{-\Delta(\Theta(d_{nm}-d))}}{1+e^{-\Delta(\Theta(d_{nm}-d))}}.
\end{equation}

Similarly as before, we link  $\Delta$ with the restriction vía 
\begin{eqnarray}
0&=&-\frac{\partial \log{(\mathcal{Z}^{(s)})}}{\partial \Delta}\nonumber\\
&=&\sum_{n<m} \frac{e^{-\Delta (\Theta(d_{nm}-d))} }{1+ e^{-\Delta (\Theta(d_{nm}-d))} }(\Theta(d_{nm}-d)), 
\end{eqnarray}
which is fulfilled only if $\Delta\to\infty$ as it needs that the sum over all non-neighbors yields zero. In that case, $P_{nm}^{(s)}=\frac{1}{2}(1-\Theta(d_{nm}-d))$ for every neighboring pair $(n.m)$ in the layer and $P_{nm}^{(s)}=0$ otherwise.  Following Eq.(\ref{A:MultProb}), the link probability for our multiplex (and thus our null model for the granular network) is
\begin{eqnarray}
P_{nm}^{(p)}&=&P_{nm}^{(s)}\times P_{nm}^{(o)}\nonumber \\
&=&\frac{\beta  \langle |Q^{\star}|\rangle^{2}}{1+ \beta \langle |Q^{\star}|\rangle^{2}}\frac{(1-\Theta(d_{nm}-d))}{2}, 
\end{eqnarray}
which can be approximated to $\beta  \langle |Q^{\star}|\rangle^{2} (1-\Theta(d_{nm}-d))/2$ for $\beta\ll1$. In this expression the term $(1-\Theta(d_{nm}-d))$ can be understood as the spatial adjacency matrix $A_{nm}^{(s)}$ for particles that are in contact (nearest neighbors). An important consequence of this approach is the entropic justification for the use of a geographic model proposed in Ref.~\cite{Daniels2018} to find community structure in a granular system subjected to simple local and global restrictions. 

\begin{figure*}[t]
\includegraphics[width=0.32\textwidth]{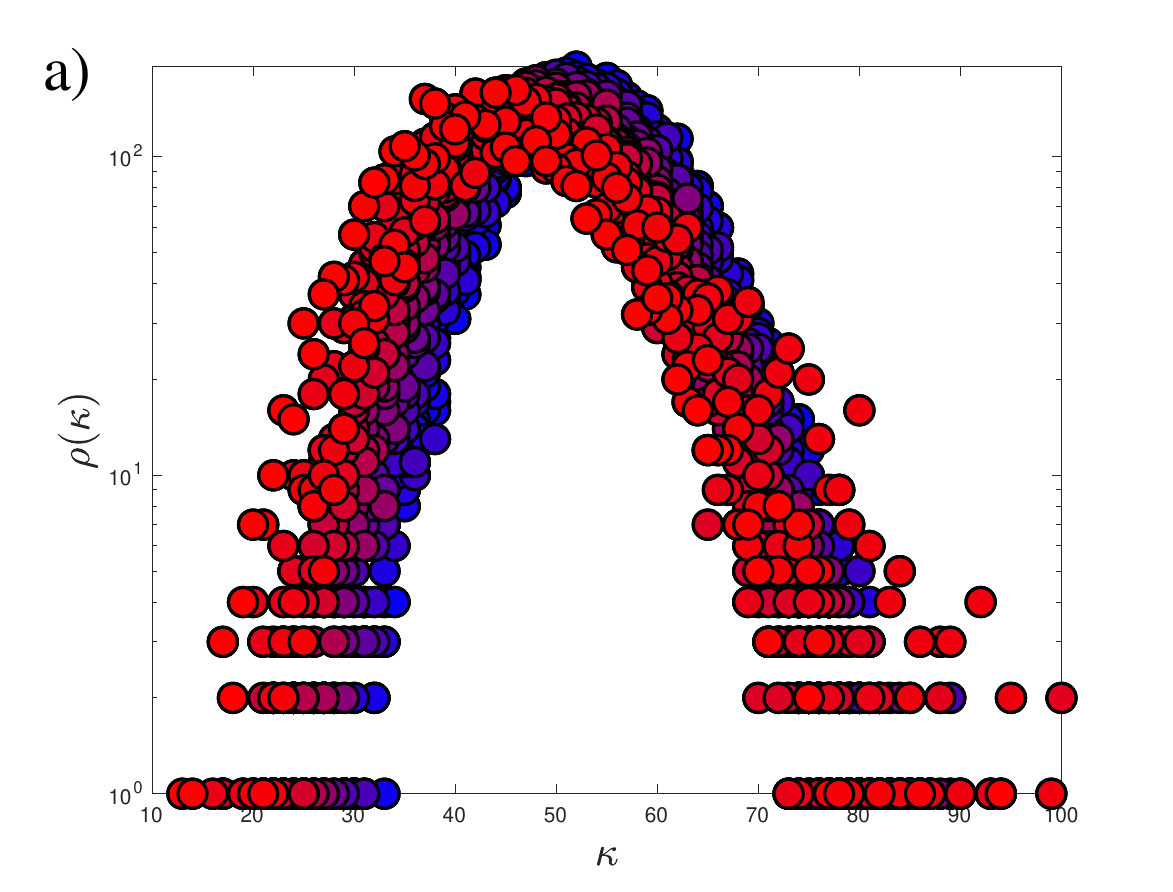}
\includegraphics[width=0.32\textwidth]{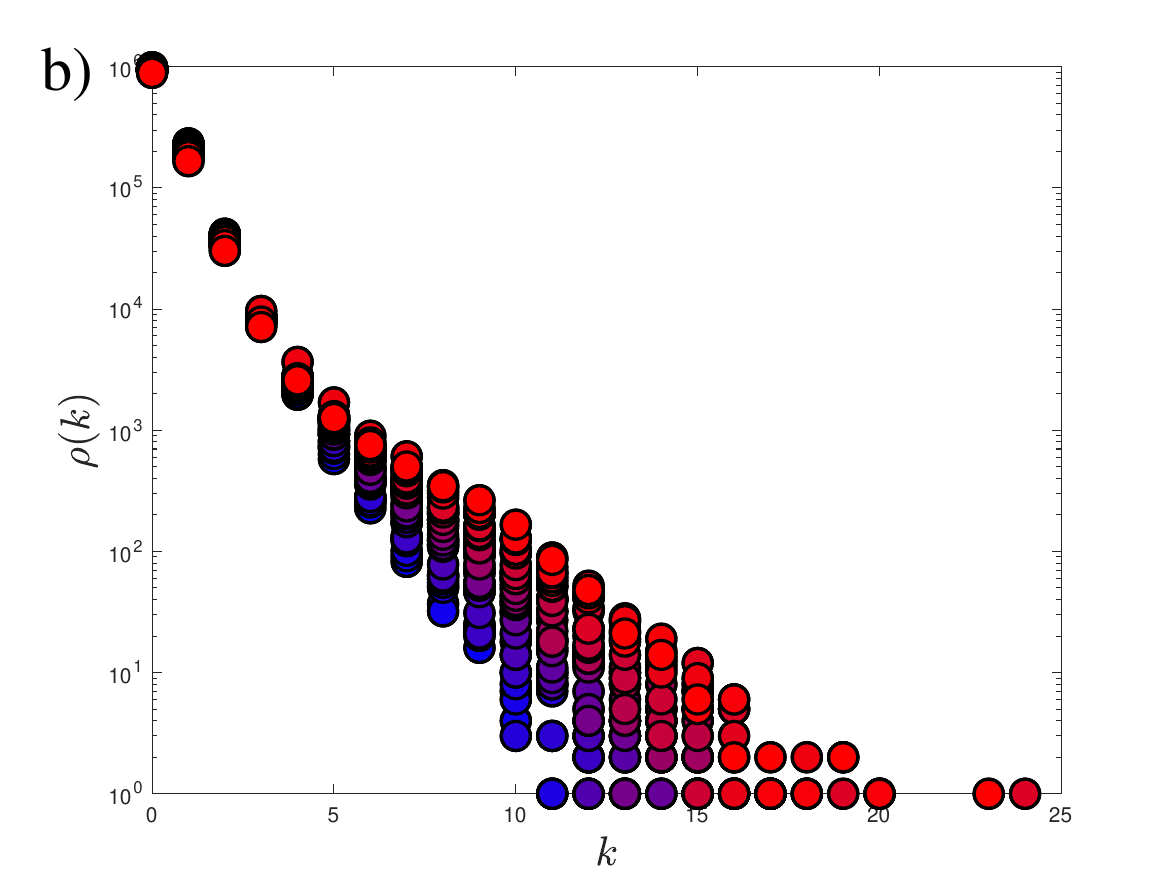}
\includegraphics[width=0.32\textwidth]{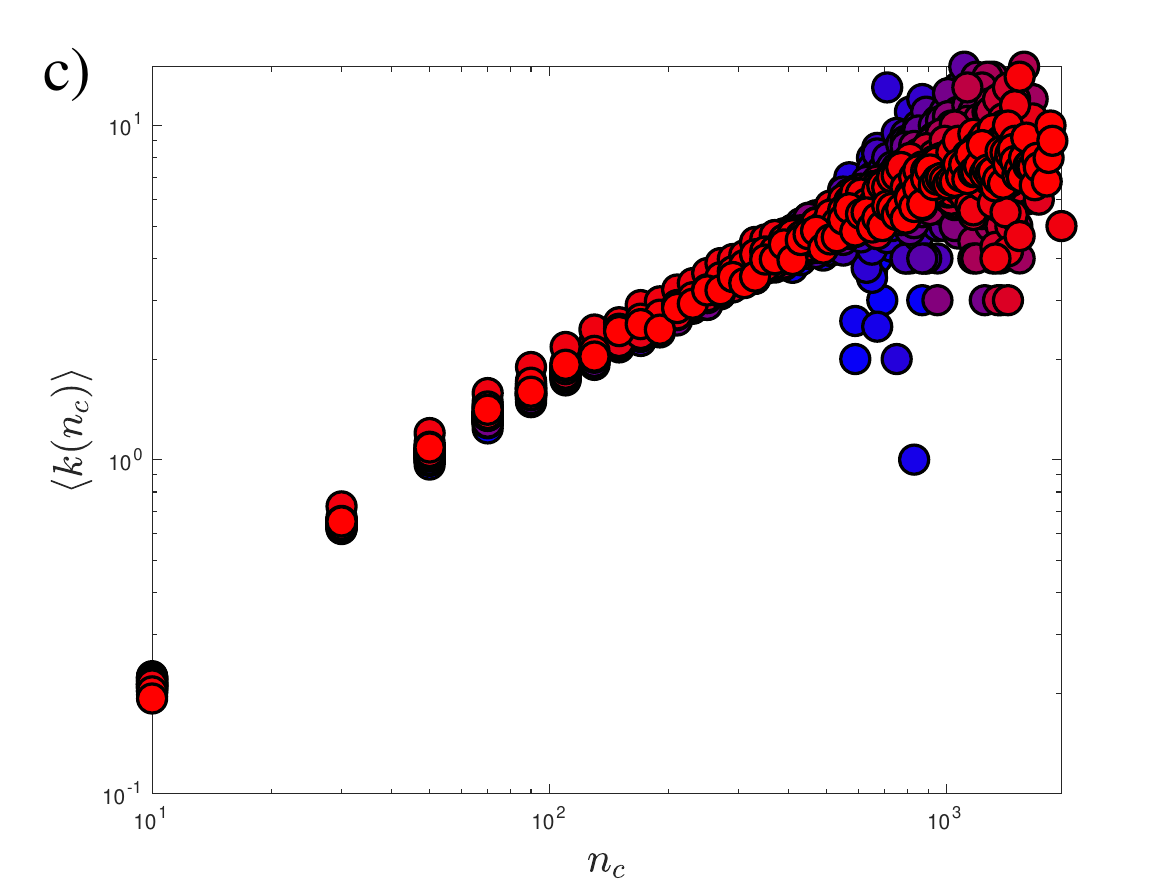}
\caption{Cluster network connection statistics. a) Distribution of the number of connections for the entire network $\rho(\kappa)$ versus $\kappa$ as a function of $\epsilon$. b)  Distribution of connections per node $\rho(k)$ versus $k$ as a function of $\epsilon$. c) Mean value of the number of connections per node $\langle k(n_c)\rangle$ versus $n_c$ as a function of epsilon. Same colorbar as in previous figures.}
\label{Fig:ClusterProp} 
\end{figure*}

\begin{figure}
\includegraphics[width=1\columnwidth]{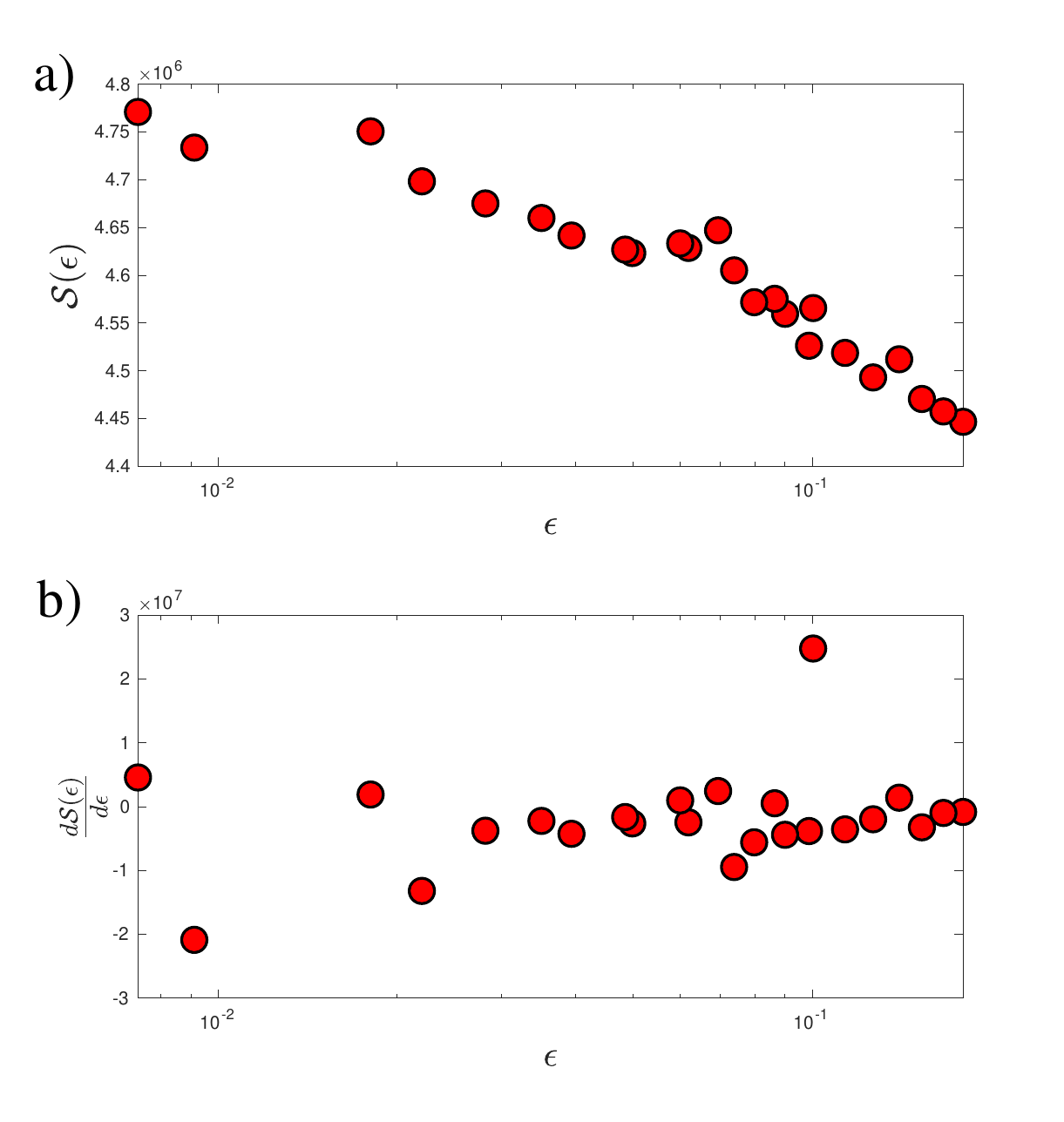}
\caption{a) Semilogarithmic horizontal plot of $\mathcal{S}(\epsilon)$ vs $\epsilon$ for the granular network. b)  Semilogarithmic horizontal plot of $\frac{d\mathcal{S}(\epsilon)}{d\epsilon}$ vs $\epsilon$ for the granular network.}
\label{Fig:Entropy}
\end{figure}

Furthermore, using these techniques, one can compute directly the entropy of the granular network via the application Eq.(\ref{A:Entropy}) to the problem at hand. As the partition function for the problem in the limit of negligible overlap is $\mathcal{Z}=\mathcal{Z}^{(o)}\times \mathcal{Z}^{(s)}$, $\mathcal{S}=\mathcal{S}(\epsilon)=\mathcal{S}^{(o)}+\mathcal{S}^{(e)}=-\Lambda L^{(o)} +\log{(\mathcal{Z}^{(o)})}+\log{(\mathcal{Z}^{(e)})}$, which can be used to compute the entropy for the ensemble of granular networks as a function of $\epsilon$, which is depicted in Fig.~\ref{Fig:Entropy}. As $\epsilon$ is the normalized acceleration which injects energy into the granular system, one could use the calculation of the granular network entropy $\mathcal{S}(\epsilon)$ as a proxy to the an out-of-equilibrium specific heat $\frac{d\mathcal{S}}{d\epsilon}\simeq C_V(\epsilon)$ which shows no divergence as $\epsilon$ approaches zero close to the solid-liquid type transition. This is consistent with the $\alpha$=0 exponent (cf. Fig.~\ref{Fig:Entropy}b)) in the language of dynamical phase transitions~\cite{HohenbergRev} within the $C-$model used to describe this granular system~\cite{Castillo2012}.

\section{Cluster network properties}

In this Appendix, we will focus on the statistical properties of the cluster network constructed above. We characterize these networks as a function of the normalized acceleration $\epsilon=(\Gamma_c-\Gamma)/\Gamma_c$. We will first deal with the statistics of the edges within the network. In Fig.~\ref{Fig:ClusterProp}a) we depict the distribution of the number of connections $\kappa$ as a function of $\epsilon$. The mean number of connections $\langle \kappa\rangle$  decreases linearly with $\epsilon$, which is expected as smaller clusters merge into larger ones, and thus less connections are formed within the network. The distribution can be reasonable fitted by a chi-squared distribution where the number of degrees of freedom $\chi$ increases with decreasing $\epsilon$.  For the case of connections per node $k$ (cf. Fig.~\ref{Fig:ClusterProp}b)), the distribution can be reasonable fitted with an exponential function for $k>4$, with a slope which decreases with $\epsilon$, allowing larger values for $k$ as we approach the solid-liquid transition. We also computed the amount of connections as a function of $n_c$, $k=k(n_c)$ grows as a power law with $n_c$ as $n_c^{\zeta_k}$ with $\zeta_k=0.55\pm0.05$ and is independent of $\epsilon$. This scaling, shown in Fig.~\ref{Fig:ClusterProp}c), is to be expected as the connections between clusters depends on the number of particles at the edges of the clusters, and thus $k\sim n_c^{1/2}$.

\end{document}